\documentclass[conference]{IEEEtran}
\usepackage[sort,nocompress]{cite}
\usepackage{amsmath,amssymb,amsfonts}
\usepackage{algorithmic}
\usepackage{graphicx}
\usepackage{textcomp}
\usepackage{xcolor}
\usepackage[hyphens]{url}

\usepackage{fancyhdr}
\usepackage[normalem]{ulem}
\usepackage[keeplastbox]{flushend}
\usepackage{subfiles}
\usepackage{subfig}
\usepackage{tikz}
\usepackage{pgfplots}

\usepackage{multirow}

\usepackage[bookmarks=true,breaklinks=true,colorlinks,citecolor=blue,linkcolor=blue,urlcolor=blue]{hyperref}

\def\BibTeX{{\rm B\kern-.05em{\sc i\kern-.025em b}\kern-.08em
    T\kern-.1667em\lower.7ex\hbox{E}\kern-.125emX}}

\makeatletter
\newcommand\resetstackedplots{
\makeatletter
\pgfplots@stacked@isfirstplottrue
\makeatother
\addplot [forget plot,draw=none] coordinates{(0,0) (1,0) (2,0) (3,0) (4,0) (5,0) (6,0) (7,0) (8,0) (9,0) (10,0) (11,0)};
}

\newcommand{\cp}{\texttt{CP}}
\newcommand{\thr}{\texttt{THR}}
\newcommand{\xor}{\texttt{XOR}}
\newcommand{\nand}{\texttt{NAND}}
\newcommand{\nor}{\texttt{NOR}}
\newcommand{\nortwo}{\texttt{NOR$_\texttt{22}$}}
\newcommand{\norone}{\texttt{NOR$_\texttt{21}$}}

\newcommand{\nortwon}{$\texttt{NOR}^\texttt{n}_\texttt{22}$}

\newcommand{\hamming}{ECiM}
\newcommand{\tmr}{TRiM}

\newcommand{\vnor}{$V_\texttt{NOR}$}
\newcommand{\vbias}{$V_{bias}$}
\newcommand{\rlow}{$R_\text{low}$}
\newcommand{\rhigh}{$R_\text{high}$}

\title{On Error Correction for Nonvolatile Processing-In-Memory}
\author{\IEEEauthorblockN{H\"{u}srev C{\i}lasun, Salonik Resch, Zamshed I. Chowdhury, Masoud Zabihi,\\ Yang Lv, Brandon Zink, Jian-Ping Wang, Sachin S. Sapatnekar, Ulya R. Karpuzcu} \IEEEauthorblockA{\textit{University of Minnesota, Twin Cities}\\\{cilas001, resc0059, chowh005, zabih003, lvxxx057, zinkx030, jpwang, sachin, ukarpuzc\}@umn.edu }
}

\begin{document}

\thispagestyle{plain}
\pagestyle{plain}

\maketitle

\noindent \begin{abstract}
Processing in memory (PiM) represents a promising computing paradigm to enhance performance of numerous data-intensive applications. Variants performing computing directly in emerging nonvolatile memories can deliver very high energy efficiency. PiM architectures directly inherit the vulnerabilities of the underlying memory substrates, but they also are subject to errors due to the computation in place. Numerous well-established error correcting codes (ECC) for memory exist, and are also considered in the PiM context, however, they typically ignore errors that occur throughout computation. In this paper we revisit the error correction design space for {nonvolatile} PiM, considering {\em both} storage/memory and computation-induced errors, surveying several self-checking and homomorphic approaches.
We propose several solutions and analyze their complex performance-area-coverage trade-off, using three representative nonvolatile PiM technologies. All of these solutions guarantee single error correction for both, bulk bitwise computations and ordinary memory/storage errors. 
\end{abstract}

\def\thefootnote{}\footnotetext{This work was in part supported by Cisco fellowships.}\def\thefootnote{\arabic{footnote}}

\section{Introduction}\label{sec:introduction}
\noindent Processing in memory (PiM) is a promising computing paradigm for
data-intensive applications.  
The core idea is performing logic operations directly within the memory system to minimize, if not eliminate, lengthy and power hungry data transfers. PiM features a rich design space spanned by (i) the underlying memory technology, (ii) where in the memory hierarchy the computation takes place, and (iii) how logic operations are performed. More traditional PiM architectures based on SRAM~\cite{aga2017compute}, DRAM~\cite{seshadri2017ambit}, {STT-MRAM} or {ReRAM~\cite{li2016pinatubo}} perform computation at the memory array periphery by exploiting sense amplifiers or by using dedicated logic blocks. In this case, the result of each logic operation has to be written back to the corresponding memory element. Computing directly within the memory arrays is also possible. This time, memory elements get seamlessly updated with the results of computation, in-situ, using different memory technologies such as DRAM~\cite{gao2019computedram,xin2020elp2im}, ReRAM~\cite{kvatinsky2014magic,imani2017mPiM}, STT-MRAM~\cite{wang2015general} or SOT/SHE-MRAM \cite{wang2021computational}.
{\em In this paper we target an especially promising class of PiM for energy efficiency, which can perform 
universal Boolean computation 
directly in nonvolatile memory (NVM)~\cite{kvatinsky2014magic,imani2017mPiM,resch2020mouse}.
}

By construction, PiM architectures directly inherit the reliability characteristics of the underlying memory substrates, but they are also vulnerable to errors due to computation.
Unfortunately, error detection and correction for PiM is not well-characterized. 
Traditional memory systems typically utilize error correcting codes (ECC) to detect and correct {storage/memory-induced} errors, 
which are not designed to protect dynamically changing data (i.e., {against computation-induced errors which PiM implies).
Adapting conventional fault tolerance techniques to protect computation is also possible, but only if
PiM logic blocks in charge of computation and the memory arrays represent separate entities, which is not always the case. 
{\em Targeted PiM architectures in this paper fuse logic and memory, where each memory cell can directly act as an input or as an output to a Boolean operation, and where computation strictly happens within the array. In this case, neither classical ECCs for storage/memory, nor classical fault tolerance techniques for computation directly apply and represent a comprehensive solution. 
We focus on this more challenging problem.
}

Numerous well-established ECCs for memory~\cite{chen1984error}, including NVM~\cite{micheloni2008error}, exist. 
A few studies consider ECCs in the PiM context:
One example 
extends MAGIC based processing in (resistive) memory
to support two dimensional 
parity bits, which enables error detection and correction in idle data only, excluding computation-induced errors \cite{leitersdorf2021efficient, kvatinsky2021making}. 
Another example covers
Triple-Modular Redundancy (TMR) \cite{leitersdorf2021making} for MAGIC based PiM in ReRAM. Redundancy here comes in two flavors: time and space.  
TMR (or generalized N-modular redundancy) is trivially simple and covers computation-induced errors,
but can incur
a significant time and/or space overhead. 

In this paper we make the distinction between conventional {\em memory} errors, that PiM systems inherit from the underlying memory, and {\em logic} errors, that stem from computing in memory.
Logic errors do not necessarily always manifest themselves as memory errors, especially when computing continuously in memory without any interruption. In this case, corruptions due to logic errors can easily propagate before periodic ECC checks to catch conventional memory errors kick in.  
Accordingly,
 we revisit the error correction design space for nonvolatile PiM, considering {\em both} memory and 
 logic errors. 
}
To this end, we explore 
classical self-checking and homomorphic approaches, and  
 introduce several solutions which can guarantee single error correction for
three representative
 nonvolatile PiM technologies supporting in-array computing semantics.  
 
The paper is organized as follows: Section~\ref{sec:background} covers PiM and fault tolerance basics; 
Section~\ref{sec:design_space_exploration}, 
ECC design space for PiM; Section~\ref{sec:detection_correction}, practical error correction approaches for PiM;
Sections~\ref{sec:experimental_setup} and~\ref{sec:evaluation},
the evaluation; Section \ref{sec:related_work}, the related work; and Section \ref{sec:conclusion}, a summary of our findings.

\section{Background}\label{sec:background}
\begin{figure*}[h]
    \centering
\includegraphics[width=.6\linewidth, trim={0cm 0cm 0cm 0cm}, clip]{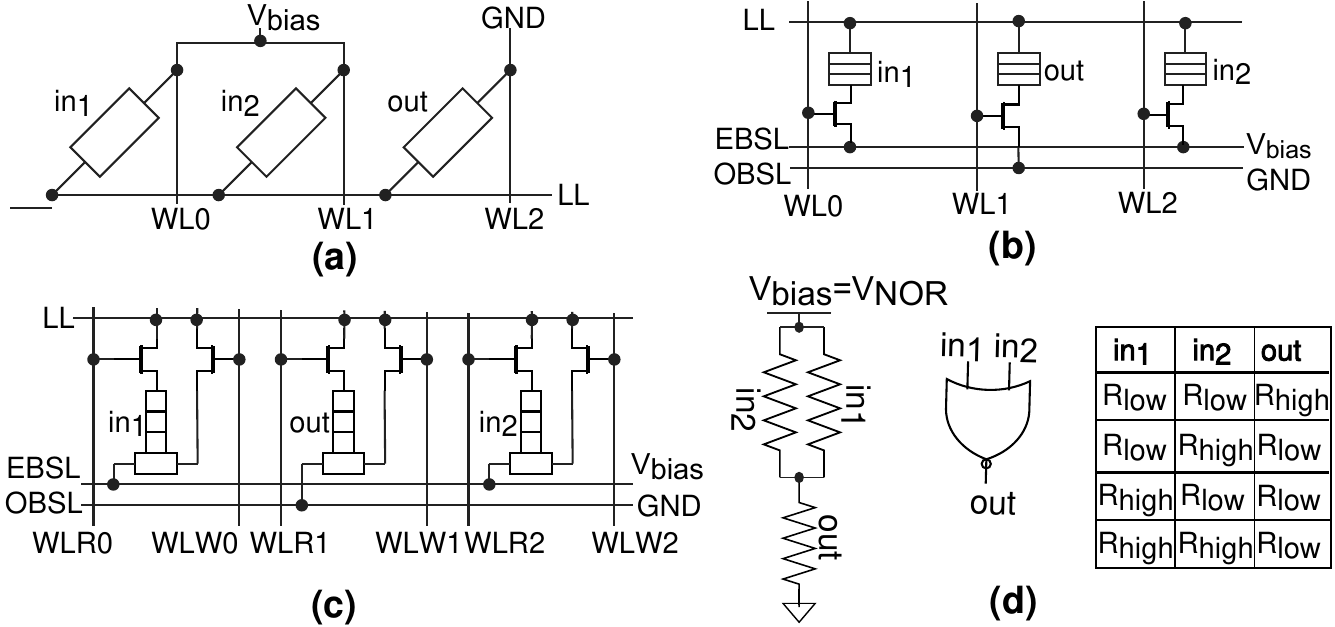}
    \caption{Logic gate 
    construction in a row of \textbf{(a)} ReRAM~\cite{kvatinsky2014magic}, \textbf{(b)} STT-MRAM~\cite{chowdhury2017efficient}, \textbf{(c)} SOT/SHE-MRAM~\cite{zabihi2018memory} arrays; \textbf{(d)} Electrical equivalent, circuit symbol and
    truth table in terms of input and output resistance levels (for STT/SHE-MRAM as a representative example) for 2-input \nor.
    \vbias\ is a gate-specific voltage applied to Bit Select Lines (BSL). (b) and (c) make a distinction between even (EBSL) and odd (OBSL) BSLs. (c) also distinguishes between Word Lines (WL) for read (WLR) and writes (WLW).
    } 
    \label{fig:array}
    \vspace{-.5cm}
\end{figure*}

\subsection{Nonvolatile PiM Basics}

\noindent  Without loss of generality, in this paper we evaluate nonvolatile (resistive)
PiM architectures which strictly perform logic operations within the memory array.  Computing directly in the arrays using memory cells already challenges ECC design due to faster and more frequent data updates in place. This problem becomes 
especially 
acute
for resistive PiM architectures capable of 
performing very large numbers of 
bulk bitwise operations in parallel, energy efficiently. 
In this case, only a minimum overhead ECC solution can help. 

Resistive memory cells encode two distinct levels of device resistance, low (\rlow) and high (\rhigh), respectively, to logic values (0 and 1, in STT and SOT/SHE MRAM; 1 and 0, in ReRAM).  Fig.\ref{fig:array} covers three representative examples based on ReRAM (a), STT-MRAM (b), and SOT/SHE-MRAM (c), which facilitate computation directly within the memory arrays using resistive memory cells.
In each case, memory functions (i.e., reads and writes) closely follow the operation semantics of the underlying memory technology. For reads, this translates into passing a small current through the memory device (by appropriately biasing the control lines),
which gets modulated by the resistance (hence, logic state) of the device, and which in turn is captured by sense amplifiers to extract the actual logic state. For writes in ReRAM,
applying a 
voltage 
at memory device's on-(off-)threshold changes the resistance to 
low(high)/1(0).
Writes in STT- or SOT/SHE-MRAM, on the other hand, have a single current threshold and the direction of the current through the memory device determines the final state. SOT/SHE-MRAM
retains most of the operation principles of STT-MRAM except that each memory device 
features a Spin-Orbit-Torque (SOT) or Spin-Hall Effect (SHE) channel to enhance the energy efficiency of writes.

For logic operations, all three designs can form universal Boolean gates within the array where each memory cell can act as an input or as an output. Electrically, each gate corresponds to a resistive network as shown in 
Fig.\ref{fig:array}(d). 
First the designated output cell is preset to a known value. 
Then, by appropriately biasing the control lines, the network in Fig.\ref{fig:array}(d) is formed. Finally, a gate specific voltage (\vbias) is applied across the network, to enforce switching of the output cell according to the underlying truth table, as a function of the logic states of the input cells.  

As an example, the output preset value for a \nor\ gate in MRAM is logic 0.
Under any \vbias\ the combined current through the input cells that passes through the 
output cell
monotonically decreases with increasing input resistance levels.
Therefore it is possible to determine a \nor\ specific \vbias\ = \vnor\ such that the output cell switches --the 
combined current through the input cells is above the critical current, $I_{C}$, for switching the output cell-- only 
if both input cells are in \rlow\
state.  
{In ReRAM, a similar procedure applies.
}
{All of the systems we cover can support
many different universal Boolean gates or gate sets besides \nor.
In the following, we will use
\nor~based bitwise logic without loss of generality.}
\label{sec:detection_correction:xor}
Due to this thresholding principle in implementing logic operations, not all logic functions can be performed in a single step. An example is \xor\, which, as depicted in Table~\ref{tab:xor_truth_table}, can 
take
3 steps (\nor, \cp, and the thresholding gate \thr\ applied in a sequence).
$\texttt{in}_\texttt{1}$ and $\texttt{in}_\texttt{2}$ here represent the inputs; $\texttt{s}_\texttt{1}$, the output of \nor; $\texttt{s}_\texttt{2}$, a copy of the \nor\ output; and $\texttt{out}$, the actual output of \xor, which is the output of the 4-input thresholding gate \thr. 
\thr\ takes $\texttt{in}_\texttt{1}$, $\texttt{in}_\texttt{2}$, $\texttt{s}_\texttt{1}$, and $\texttt{s}_\texttt{2}$ as inputs. The preset for \thr\ output is logic 0, which only switches 
to 1 if three or more of its inputs are 0.
All of the the PiM technologies we consider support 2-output \nor\ gates, \nortwo, which 
makes finishing the
\nor\ and \cp\ gates from Table~\ref{tab:xor_truth_table} in one step possible, rendering a 2-step \xor\ function (2-output \nor\ and \thr\ performed in a sequence). In the Appendix, we provide a detailed electrical characterization for 2-output gate operation.

\begin{table}[htp]
\vspace{-.1cm}
    \centering
    \scalebox{0.97}{
    \begin{tabular}{c|c||c|c||c}
         &  & $\texttt{s}_\texttt{1}=$ & $\texttt{s}_\texttt{2}=$ & $\texttt{out}=$\\
         $\texttt{in}_\texttt{1}$& $\texttt{in}_\texttt{2}$ & $\nor(\texttt{in}_\texttt{1},\texttt{in}_\texttt{2})$ &  $\cp(\texttt{s}_\texttt{1})$ & $\thr(\texttt{in}_\texttt{1},\texttt{in}_\texttt{2},\texttt{s}_\texttt{1},\texttt{s}_\texttt{1})$\\
        \hline
        0 & 0 & 1 & 1 & 0 \\
        0 & 1 & 0 & 0 & 1 \\
        1 & 0 & 0 & 0 & 1 \\
        1 & 1 & 0 & 0 & 0
    \end{tabular}
    }
    \caption{3-step \xor~\cite{chowdhury2019spintronic}.}
    \label{tab:xor_truth_table}
    \vspace{-.2cm}
\end{table}

All three PiM technologies support three different levels of parallelism: (1) Partition-level  parallelism \cite{leitersdorf2022partitionpim} where each row can be divided into several partitions using transistors, such that multiple logic operations can be performed in each row; (2) Row-level parallelism, where each row can perform the same Boolean gate simultaneously;  (3) Array-level parallelism where each memory array can perform 
computational tasks in parallel.

{

\subsection{Application Mapping}
\noindent 
A generic PiM compiler flow incorporates
three major steps as sketched in \cite{resch2020mouse}:

\begin{list}{\labelitemi}{\leftmargin=.5em}
  \itemsep0.5em 
    \item[{\bf 1)}] \textbf{Intermediate code generation} 
    involves identifying (multi-bit) PiM operations as well as the data layout for the input, output, and scratch spaces, {i.e.}, cells where the inputs and the outputs reside and where the intermediate gate operations take place, respectively. 
    \item[{\bf 2)}] \textbf{Gate-level opcode generation} 
    translates multi-bit operations from 1) 
    into actual Boolean logic gates from the PiM library.
    HDL synthesis tools can be adopted for this level of translation.
    Also possible is taking advantage of emerging
    homomorphic computing tools, where software transpilers can map arbitrary software to Boolean gates \cite{hallman2018building,carpov2015armadillo,chielle2018e3,lee2020optimizing,gorantala2021general}.
    \item[{\bf 3)}] \textbf{Binary instruction translation} represents the last step, where gate-level opcodes get mapped to their binary equivalents to drive voltages. This mapping is specific to the array architecture and involves the bias voltages for BSLs/WLs from Fig.\ref{fig:array}.
\end{list}

This broad classification can take different shapes depending 
on the location of the underlying PiM substrate in the memory hierarchy. 
If, e.g., the PiM substrate represents
a custom accelerator, 3) or both 2) and 3) can be 
undertaken by the accelerator.
Otherwise, the memory controller of the respective level in the memory hierarchy can 
perform the translation in 3), while the rest can be handled purely in software.
}

\subsection{Fault Tolerance} \label{sec:background:fault_tolerance}

\noindent{\bf Errors} 
in a computing system can broadly be classified as \emph{hard} (permanent) and {\em soft} (temporary) errors.
Previous work in the PiM context further distinguishes between soft errors induced by intended operations such as a faulty write or an incorrect logic operation (referred to as {\em direct} errors) and other soft errors 
(referred to as {\em indirect} errors)~\cite{leitersdorf2021making}. Direct errors form the focus of our study.
Bulk bitwise logic operations in NVM can be subject to different reliability issues. For example, memristive devices with voltage dependent switching may suffer from resistance fluctuations~\cite{zhu2020implication}, or the switching of spintronic devices may probabilistically change due to the initial magnetization or thermal fluctuations~\cite{liu2014dynamics}.
In case of ReRAM another major error 
source is the resistance state confusion \cite{long2019design},
which can be quantitatively characterized by the  
overlap between high and low resistance regions. 
More precisely, 
ReRAM is vulnerable to diffusion of oxygen vacancies, ion strikes or environmental factors, in a both temporally correlated and uncorrelated fashion \cite{leitersdorf2021efficient}. MRAM is prone to thermal noise \cite{hoffer2022performing}, retention failures, read disturbance, write errors, 
write pulse/current variations \cite{motaman2015impact,sun2012process}, tunneling magnetoresistance ratio variations \cite{zhao2012failure}, and bias voltage variations \cite{zhang2011stt}. 
For key parameters such as MRAM tunneling magnetoresistance ratio \cite{hoffer2022performing} or ReRAM threshold voltage \cite{li2022error}, Gaussian distributions apply. 
Regardless of the origin, however, such nonidealities mainly manifest themselves as single bit flips for STT-MRAM \cite{lv2023experimental}, SOT-MRAM \cite{hoffer2022performing}, or ReRAM \cite{li2022error}.
{For gate operations, such single bit flips stem from the output not being able to change state when it is supposed to, or vice versa. } 
As these technologies are not mature enough, computation-induced errors in PiM are not well characterized.
We also should note that 
to be feasible for practical applications, respective gate error rates should 
not significantly exceed
the error rates of conventional memory technologies of today. 

Aside from technology specific adjustments, our error model is based on previous work~\cite{leitersdorf2021efficient,leitersdorf2021making}, where, without loss of generality, errors in Boolean gate operations are  uniformly distributed in each PiM array throughout row-parallel computation.

\noindent {\textbf{Self-checking Circuits} 
have the ability to 
discern incorrect operation by monitoring their own outputs \cite{anderson1971design, wakerly1974partially}. 
This is enabled by adding redundancy in the form of check symbols to the data representation.  
Two types of self-checking circuits exist: Type-I features \emph{systematic}; Type-II, \emph{non-systematic} codes.  
In {systematic} codes,
the actual data and check symbols exist as separate distinct components
in the overall {codeword}. Here, direct access to the data to be protected is possible. However, 
error vulnerability of 
the check symbols
makes it difficult to 
distinguish between errors in the actual data vs. errors in the check symbols.
In {non-systematic codes},
actual data and check symbols are combined in an interleaved fashion to form
{codeword}s. The obvious disadvantage is lack of 
direct access to 
actual data, 
while 
error assessment becomes easier.}

\noindent{\bf Modular Redundancy}
in broad terms
entails 
using multiple copies of 
unreliable components to 
improve reliability. In Dual Modular Redundancy (DMR), two copies of the exact same computational primitive are executed
in parallel or series, the outputs are compared, and an error is signaled in case of a mismatch.
DMR can detect but not correct errors.  Only if the outputs match the computation is considered correct. Hence, for DMR to work, probability of two simultaneous errors (one in each copy) should be lower than the probability of a single error. 
Similarly, Triple  Modular Redundancy (TMR) relies on three copies where the computation is deemed correct if a strict majority of the outputs match. For TMR {to} work, two simultaneous errors (one in each copy) should be less likely than a single error, as well. TMR, however, can correct 
up to 
one error. 
In the context of PiM operations, N-modular redundancy translates into performing N copies of the respective gate operation, which necessitates working with N copies of the corresponding input operands, as well. Targeted PiM architectures in this paper are promising due to the capability of performing bulk bitwise operations in a massively parallel fashion, and the flexibility of being able to fire computation in any row within the memory array. Hence, even for modest N (which is the case for DMR or TMR), {the area and/or time} overhead can easily become prohibiting.

\noindent{\bf Hamming Codes} \cite{hamming1950error}
are linear block codes\footnote{Linear combination of the codewords is another codeword.  Block codes can be applied on chunks of limited size data independently.} featuring 
a Hamming distance of 3 between any two codewords, therefore, can {correct (detect)} single (double) bit errors.
In formal terms, $n$ bit Hamming codewords are obtained by multiplying each $k$ bit binary vector corresponding to raw data by a binary generator matrix {$G_{k{\times}n}$}. 
Error checks translate into multiplying another binary matrix {$H_{(n-k){\times}n}$} termed parity-check matrix by codewords, where
\begin{equation} \label{eq:hamming}
G=\{I_k|-A^T\} \qquad H=\{A|I_{{n-k}}\}
\end{equation}

\noindent applies. 
$A$ here is a {$(n-k){\times}k$} binary submatrix; $I_k$, the $k \times k$ identity matrix; $T$, the transpose operation; and $|$, {horizontal} concatenation. 
{$A$ provides a mapping from raw data to codewords
such that information in redundant bits
can 
pinpoint erroneous bit flip locations.}
Considering matrix dimensions, multiplying by $G$ extends the $k$ bit raw data vector by $n-k$ bits.
During check, the $n$ bit codeword 
is multiplied by $H$, which produces an $n-k$ bit vector called the \emph{syndrome}. If the syndrome is all zero, there is no error. Otherwise, each possible syndrome value points to a unique location for an error in the actual (raw) data vector, which in turn can be 
corrected by a simple bit flip.
Codeword generation (encoding) and syndrome computation translate into modulo-2 matrix multiplications.
{
Hence, a sequence of \texttt{AND}s suffice for element-wise multiplication, where modulo-2 addition
reduces to \xor.}
Hamming codes can be implemented in systematic or non-systematic form; and {conversion between the two forms takes elementary matrix transformations.} While the  coverage is similar to classical TMR,  
number of check bits grow as $\log{(n+1)}$ with increasing code word length $n$.

\section{ECC Design Space}\label{sec:design_space_exploration}
\begin{figure}[th]
\centering
\subfloat[Column-wise]{\label{fig:ecc_columnwise}{
\includegraphics[width=.7\linewidth,trim={0cm 14.5cm 23.8cm 0cm},clip]{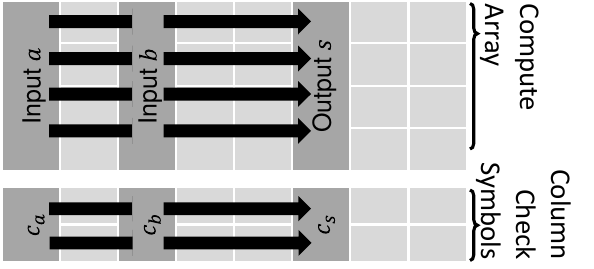}}}\\
\vspace{-.2cm}
\subfloat[Row-wise]{\label{fig:ecc_rowwise}{\includegraphics[width=.75\linewidth,trim={0cm 15.5cm 22.8cm 0cm},clip]{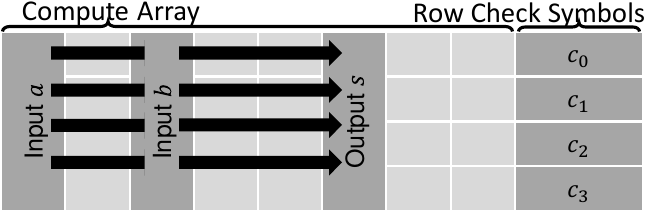}}}
\caption{
Check symbol layout. 
}
\label{fig:ecc_classes}
\vspace{-.5cm}
\end{figure}

\noindent We can envision each ECC codeword as 
a combination of raw data and \emph{check symbols}, i.e., the accompanying redundant information that enables error detection and/or correction.
Check symbols in a codeword can be totally isolated from the raw data bits ({systematic} ECCs); or interleaved ({non-systematic} ECCs). In the following, we stick to systematic ECCs where data to be protected can be accessed directly, which by construction enables a more modular design, especially useful in the PiM context. 
In this case, as computation is performed in each row, the check symbols can be organized either column-wise (Fig.\ref{fig:ecc_columnwise}) or row-wise (Fig.\ref{fig:ecc_rowwise}).
In this illustrative example, all rows compute in parallel, processing a single bit of the inputs $a$ and $b$ to generate a single bit of the output $s$.

\subsection{Column-wise 
ECC} \label{sec:concurrent_ed}
\noindent In Fig.\ref{fig:ecc_columnwise}, a dedicated portion of the memory array 
is allocated for computation, where bulk bitwise logic operations are performed.
Each column has its check symbols in a separate set of rows, which may form an isolated portion in the same memory array or a separate array.
Most significantly, column check symbols of the output column are calculated using the column check symbols of the input columns, which can proceed simultaneously with the actual computation. Deriving the output column check symbols from the input check symbols in this manner, however, is only 
practical
if the following three criteria are satisfied:

\begin{list}{\labelitemi}{\leftmargin=.5em}
\itemsep 0.5em 
    \item[1)] The bitwise operation in Fig.\ref{fig:ecc_columnwise} is generic.
    We can assume that it corresponds to any of the universal Boolean gate operations \nand\ or \nor, where $s={\bar{a} \lor \bar{b}}$ or $s={\bar{a} \land \bar{b}}$ applies, respectively. 
    In this case, the check symbols should satisfy
    ${s = \bar{a} \lor \bar{b}} \longleftrightarrow c_s = f(c_a, c_b)$ or ${s = \bar{a} \land \bar{b}} \longleftrightarrow c_s = f(c_a, c_b)$, 
    where $c_a$ and $c_b$ correspond to the check symbols for $a$ and $b$; $\longleftrightarrow$ depicts logic equivalence; 
    and $f$ is an appropriately defined ECC operator that generates $c_s$, i.e., the check symbols for the output of the 
    \nand\ or \nor.
    $f(c_a, c_b)$ in this case only depends on $c_a$ and $c_b$ (and not on actual data). This enables ECC updates using check symbols only.
    However, not all ECCs can satisfy this criterion.
    \item[2)] The check symbols $c_a, c_b$ should have a modest storage requirement compared to the actual (raw) data.
    \item[3)] Provided that 1) applies, as the main contributor to the ECC overhead, calculation of $f(c_a, c_b)$ should be computationally cheap.
\end{list}

The first 
criterion directly restricts the type of applicable ECCs. This criterion essentially is after homomorphic operation, which guarantees that computation on raw data can always be mapped to computation on check symbols without any ambiguity. 
Under typical homomorphism, bitwise logic operations on raw data map to element-wise addition and multiplication between very long codewords, which often translates into very complex numerical operations. It is hard to justify the computational overhead, especially for bulk bitwise operations, where a single Boolean logic gate is of concern.
As a result, promising homomorphic linear block codes such as Reed-Muller~\cite{cho2020homomorphic} satisfy 1), but not at all 2) and 3). This generally applies to other candidates including  homomorphic arithmetic codes such as Berger codes~\cite{lo1992sfs}. 
{\emph{
Typical homomorphic codes fall short of satisfying 2) and 3) for bulk-bitwise logic in PiM, rendering column-wise 
ECC infeasible.
}}

\subsection{Row-wise ECC}
\noindent The alternative 
is keeping check symbols on a per row basis, as depicted in Fig.\ref{fig:ecc_rowwise}.
In this case, each operation on the compute array requires an update on the row check symbols, which,
when compared to the column-wise alternative, can incur a higher time overhead (especially if only one gate operation can be performed in each row at a time). 
However, even under partition-level parallelism (where each row is chunked into multiple sub-rows 
and where a gate operation can be performed in each sub-row simultaneously)  check symbol updates cannot be performed {\em fully} parallel to logic operations on actual data bits, in stark contrast to column-wise schemes. This is because row check symbols $c_0$, $c_1$, $c_2$, $c_3$ each represent a function of one bit of the output $s$, and hence, cannot be computed without the corresponding bit of the output being ready.
On the other hand, for the row-wise case 
no other restrictive criterion (such as homomorphism
for the column-wise alternative) applies.
The only practical criterion for the row-wise case is a time- and space-efficient implementation of check symbol updates.  As we are going to cover in the next section, satisfying this criterion is possible by 
overlapping 
check symbol updates and actual computation on data bits 
using Hamming codes.

{
\subsection{Putting It All Together}
\noindent Although column-wise ECC {(Section
\ref{sec:concurrent_ed})
} provides a more elegant mathematical solution, the 
overhead due to complex arithmetic remains much higher than 
row-wise 
ECC. This, however, doesn't necessarily imply that row-wise ECC is practical. 
Row-wise ECC serializes check symbol updates after actual computation, which
can incur significant time
overhead, 
easily on a par with trivial modular redundancy.
In the following, we will demonstrate to what extent overlapping actual computation with ECC updates can help address this challenge. 
}

\section{Error Correction in Nonvolatile PiM}
\label{sec:detection_correction}
\begin{figure*}[h]
    \centering    
     \includegraphics[width=.85\linewidth]{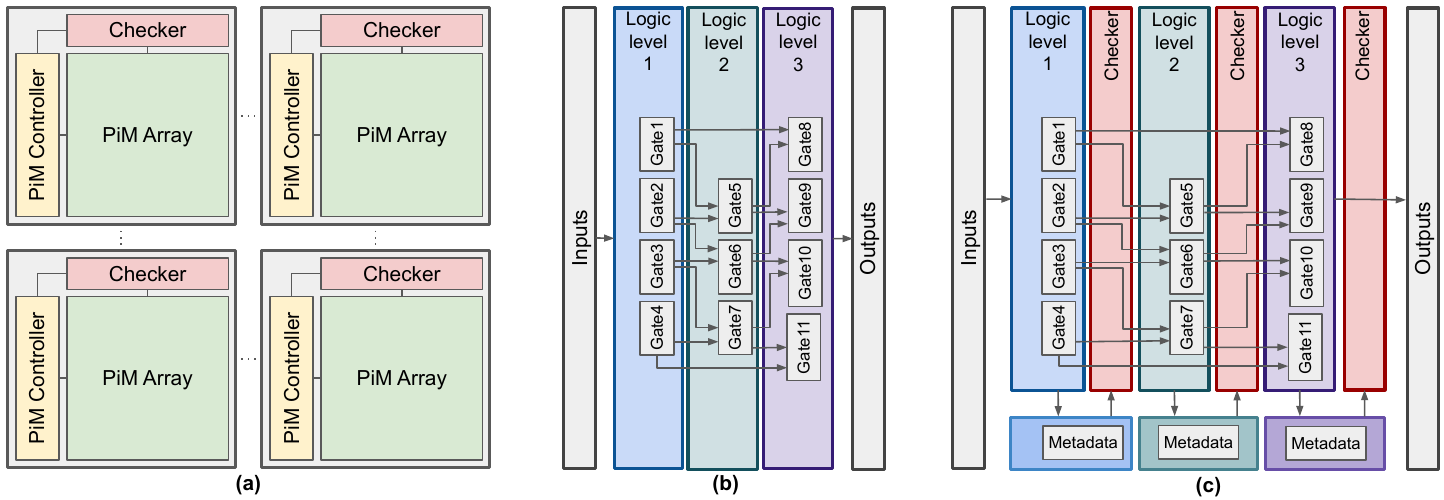}
    \caption{\textbf{(a)} Overall system architecture. \textbf{(b)} Main computation in (one row of) memory with logic levels explicitly shown. \textbf{(c)} Main computation (in one row of memory) interleaved with error detection/correction as performed by {\em Checker} blocks. For error detection, \hamming\ {\em Checker} blocks are in charge of syndrome generation; \tmr\ {\em Checker} blocks, majority vote calculation. {\em Metadata} translates into parity bits for \hamming; two redundant computation outputs, for \tmr.
    }
    \label{fig:overall_scheme}
    \vspace{-0.4cm}
\end{figure*}

\subsection{Coverage: Single Error Protection (SEP)}
\label{sec:tech_cover}
\noindent We target novel nonvolatile technologies that can perform logic gates directly in memory arrays, which are still being developed. Device level reliability has to be improved for any practical use scenario. 

For example, today, the spin-transfer torque magnetic random-access memory (STT-MRAM)~\cite{8662444,8993469,9062955,9371922} is the most mature MTJ technology, with commercial offerings from Everspin, GLOBALFOUNDRIES, IBM, Intel, Samsung, TSMC, and Avalanche Technology, albeit with great restrictions on the structure of the MTJ array. One of the chief benefits of STT-MRAM over competing nonvolatile technologies lies in its endurance, energy efficiency, and speed. In laboratory conditions, it has been reported to exhibit endurance up to $10^{14}$ write cycles~\cite{10019430}, energy consumption as low as 90fJ/bit~\cite{10.1063/1.4869828}, delay as low as less than 200ps~\cite{Zhao_2011}, and Tunneling Magnetoresistance Ratio as high as 600\%~\cite{10.1063/1.2976435}. Gate error rate is very sensitive to {Tunneling Magnetoresistance Ratio}, and can significantly decrease with modest increases in {this ratio}. Though all of these metrics may not be simultaneously achievable today (for example, a recent Samsung’s product~\cite{10019430} offers $10^{14}$ cycles endurance, 100ns write speed, and 25pJ/bit write energy consumption, where a product line by Avalanche Technology~\cite{Avalanche} offers $10^{14}$ cycles endurance, 20ns write speed, and 10pJ/bit write energy consumption), rapid technology developments 
are promising.

Accordingly, 
once these technologies mature, they will reach reliability levels similar to conventional memory technologies of today, where {TMR (Triple Modular Redundancy)} level coverage is adequate.
Accordingly, we target single error protection (SEP) in this paper.

\subsection{Full System Design: Macroscopic View}

\noindent In the following, we will explore the design space for novel PiM specific SEP solutions considering two approaches. The first approach relies on fine grain parity calculation to efficiently implement Hamming codes. The second approach rethinks ordinary TMR to protect in-memory computations. We will refer to the resulting designs as \hamming\ and \tmr\ respectively.  

The key design challenge is keeping the overhead of error detection and correction at bay. 
The end performance primarily depends on 
at what granularity in time and space error detection takes place, and
upon error detection, how/where error correction is performed:   

\begin{list}{\labelitemi}{\leftmargin=.5em}
  \itemsep0.5em 
\item{Ideally, to prevent error propagation, a check for errors should happen after each logic operation to trigger correction immediately as necessary. While the proposed \hamming\ and \tmr\ designs
can accommodate this,        
{\em to keep the SEP overhead at bay,
we perform checks 
at logic-level granularity}.
This does not necessarily compromise coverage, as Boolean operations within the same logic level are typically not data-dependent.
This way,  we can guarantee SEP in the main computation -- as well as 
parity calculations for \hamming\ and redundant operations for \tmr.
}
\item{If error correction happens directly in memory, the correction logic becomes subject to the very same errors as the main computation, and can become a bottleneck.
To address this, our designs feature
dedicated error correction blocks, hardened by design, outside of (but near) PiM arrays.}
\end{list}

\noindent {\bf Error Detection \& Correction Semantics:} Fig.\ref{fig:overall_scheme} provides the full system overview. 
{\em Checker}
blocks depict external logic blocks dedicated to error detection and correction. For \hamming, each {\em Checker} processes the parity information from a PiM array to calculate the syndrome for error detection. For \tmr, each {\em Checker} processes the output of the main computation along with the two redundant copies to calculate the majority vote for error detection. Error detection happens at logic level granularity. Each {\em Checker} block is also in charge of error correction and sending the corrected data back to the respective PiM array.

\hamming\ updates parity information in memory after each logic operation, as detailed in Section~\ref{subsec:parity_pipeline}. However, checking for errors happens at the granularity of logic levels outside each PiM array. 

Whenever the computation of all Boolean gates in a logic level along with the accompanying parity updates is complete, we transfer the result of the computation
as well as parity bits to the {\em Checker} through a conventional memory read. Each logic level occupies a single row of the PiM array.
{The first phase of the check (decoding in more formal terms) is to multiply the $H$ matrix (from Equation~\ref{eq:hamming}, which is hardwired in the {\em Checker}) with the vector (codeword) incorporating main computation results and parity bits to obtain the syndrome vector. 
Additions in matrix-vector multiplication reduce to \xor{}s; and bitwise multiplications, to \texttt{AND}s.
The next phase is correction, implemented by \xor{}ing the syndrome vector with the vector carrying main computation results.
} 
\hamming\ {\em Checker}s therefore represent relatively light-weight hardware blocks.
If an error is detected, 
the final phase is writing 
the corrected logic level output back to the PiM array. 

\tmr\ calculates two redundant copies in memory, in the same row. 
Whenever the computation of all Boolean gates in a logic level along with two redundant copies is complete, we transfer the result of the computation
as well as redundant outputs to the {\em Checker} through a conventional memory read. 
{The first phase of the check is to take the majority vote among the three copies for error detection. 
If an error is detected (as signaled by a mismatch in the outputs), the final phase is writing 
the correct logic level output (the output with the majority vote) back to the PiM array. 
{\em Checker}s in this case primarily rely on majority voting logic and small multiplexers, hence, constitute relatively light-weight hardware blocks, as well.

\begin{figure} [thp]
\vspace{-.4cm}
\centering
\subfloat[Operation layout (\hamming)]{\includegraphics[width=.65\linewidth]{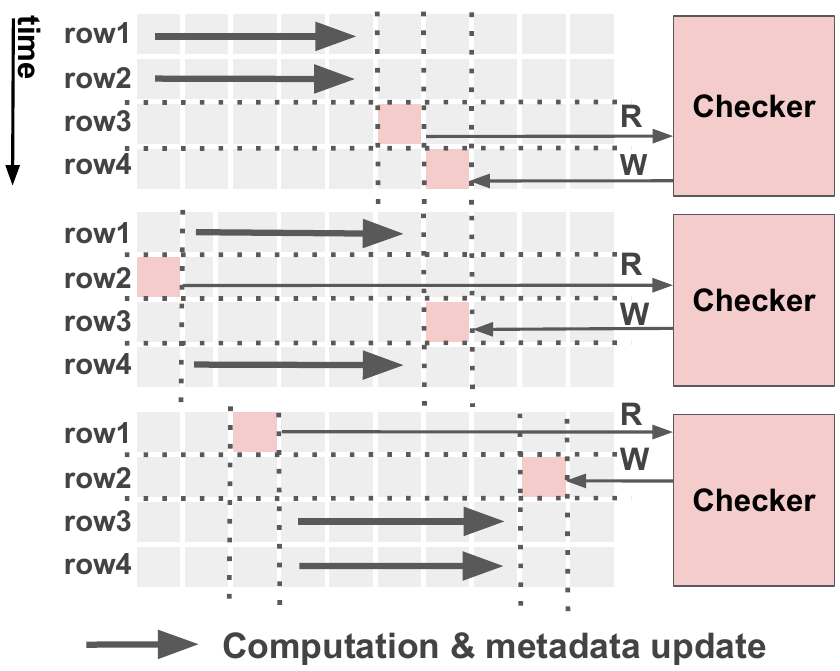}\label{fig:hamming_macropipeline_wtmr}}\\
\subfloat[\hamming\ (top) and \tmr\ (bottom) timing]{
\includegraphics[width=\linewidth,trim={0cm 0cm 0cm 0cm}, clip]{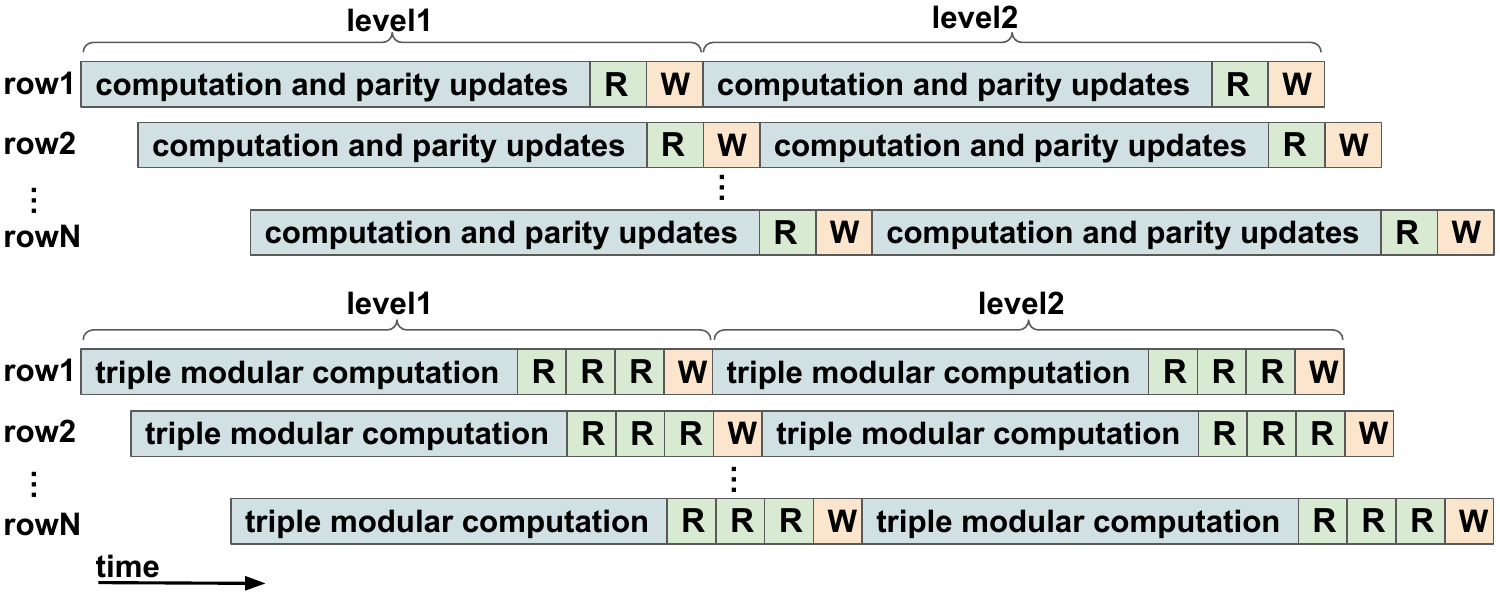}
\label{fig:hamming_timeline}}\\
\caption{
Operation layout (a) and timing (b). Each row independently computes the same gates, logic level by logic level, on different data. To overlap computations in one row with reads or writes in other rows, computations in each row start in a delayed fashion (b). More specifically,
recall that we use universal \nor\ gates as core building blocks for computation; all computations are synthesized using \nor. 
With delayed start, when $r^\text{th}$ row executes $s^\text{th}$ \nor, $(r+1)^\text{st}$ row would execute $(s-1)^\text{st}$ \nor\ of the same level, on different data. 
}
\vspace{-0.2cm}
\end{figure}

\noindent {\bf Practical Considerations:} 
If not carefully orchestrated, data communication with {\em Checker} blocks can become a performance bottleneck. 
In the PiM execution model each row operates in parallel, performing (all logic levels of) the same computation on different data (Section~\ref{sec:background}). 
We cannot initiate error checking for a row until the computations and metadata (parity for \hamming; redundant copies for \tmr) calculations in the same row are complete, at the logic level granularity. 
 However, we can interleave R(ead) and W(rite) operations in a given row (as induced by the communication with the {\em Checker}) with computations and metadata updates in other rows 
 to reduce the overall time overhead -- as demonstrated by the operation layout in Fig.\ref{fig:hamming_macropipeline_wtmr} (for \hamming, without loss of generality); and timing, in Fig.\ref{fig:hamming_timeline}.
 The goal is to fully utilize the communication bandwidth at the PiM array interface.
 To this end, we (i) partition columns, similar to rows (Section~\ref{subsec:parity_pipeline}); 
 (ii) adjust the layout of PiM operations in each row.

\begin{figure}[t]
    \centering
    \includegraphics[width=\linewidth,trim={0.05cm 0cm .05cm 0cm},clip]{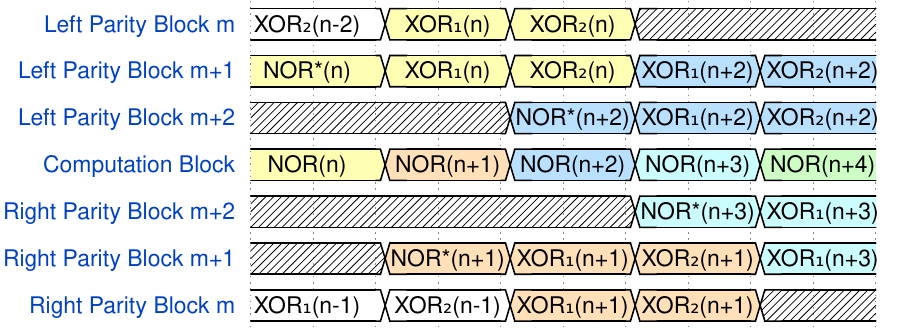}
    \caption{Timing diagram for parity updates vs. main computation.
    {
    Each waveform captures the activity in a specific block 
over 
time.
The hatch pattern depicts an idle block.
The computation (\nor) that triggered the parity update 
(i.e., the two steps of \xor) and the corresponding parity update operations are labeled using the same color. 
    {\nor}* and {\nor} in this diagram point to the very same \nor\ gate in actual computation, where{\nor}* indicates the calculation of the second output of \nortwon\ in a different block. $\texttt{XOR}_1$ and $\texttt{XOR}_2$ denote the two steps of \xor;  \nortwo\ and \thr, respectively. For each gate, the corresponding step in computation is indicated in parenthesis: {\nor}\texttt{(n+1)} is the \nor\ gate initiated at Step \texttt{n+1}. } 
    }
    \label{fig:waveform_parity}
    \vspace{-.7cm}
\end{figure}

\subsection{Metadata Updates in \hamming}
\label{subsec:parity_pipeline}

\noindent In this section we take a closer look into metadata updates. We start with how \hamming\
updates parity information after every (\nor) operation throughout computation. For \hamming, this forms the backbone for generating Hamming codes in memory.

\noindent {\bf Parity Updates in Memory:} 
Targeted PiM architectures perform computations in rows. Hence, actual computation and parity updates can happen in separate, dedicated columns, closely following row-wise error detection semantics from Fig.\ref{fig:ecc_rowwise}.
PiM technologies that we consider in this paper can implement multiple-output gates with identical outputs 
(Section~\ref{sec:background}), which
we can exploit for seamless parity updates.
To this end, for example, we can use the two-input\slash two-output \nortwo\ gate instead of a standard two-input/single output \norone\ gate. 
The first output is produced in dedicated computation columns;
the (identical) second output,
in 
dedicated parity columns. 
We keep a running parity bit in each row, per logic level.
The second output gets \xor-ed with the 
running
parity bit
to update the instantaneous parity information. 
Whenever a \nortwo\ output is one, it flips the corresponding running parity bit. 
Each parity update triggers the \xor\ of (the instantaneous value of) the respective parity bit with one of the identical outputs of the \nortwo\ (corresponding to the instantaneous step of computation). 
Each such \xor\ can be performed as a two step operation consisting of a \nortwo\ and a thresholding \thr\ gate (Section~\ref{sec:background}). 
Hence, each parity update (triggered by a \nortwo\ operation in the computation columns) 
gives rise to two extra gate operations (\nortwo\ and \thr).
Fig.\ref{fig:waveform_parity}
captures how main computation gets interleaved with parity update operations to form a pipeline:
The PiM array is partitioned 
into {\em left parity columns}, {\em compute columns}
 in the middle, and {\em right parity columns}. Parity columns keep intermediate data during parity updates.
 Specifically, left and right parity columns keep the parity after alternating steps of computation, respectively, and help pipeline parity update and actual computation operations to minimize the time overhead. 
 Moreover, left and right columns are organized as independent blocks, where each block corresponds to a separate partition. 
 {Essentially each block corresponds to a number of neighboring columns, which are separated from the rest of the columns by switches in the logic lines (LLs from Fig.\ref{fig:array}).}
 Partitioning semantics closely follow \cite{leitersdorf2021making}, by en/disabling switches in logic lines (which establish electrical connections between inputs and outputs of logic operations).
 No more than one logic gate operation can be in progress in one partition at a time, while gate operations can span (have inputs and outputs distributed over) multiple partitions as long as there are no other overlapping simultaneous gate operations in progress in any of the partitions.

To be more specific, the left and the right side each keeps a separate parity bit,
each corresponding to a sequence of alternating gate operations (which we can think of as the odd and even numbered gate operations in the compute columns, respectively, if gate operations were hypothetically enumerated in a sequence at runtime).
Every computation in the compute columns triggers a parity update of this form, on one specific side of the parity blocks.
Recall that each parity update (triggered by a \nortwo\ operation in the compute columns) 
gives rise to {an} \xor\ implemented by two extra gate operations, \nortwo\ and \thr, respectively, which we will refer to as $\texttt{XOR}_{1} = \texttt{XOR}_{\nortwo}$ and $\texttt{XOR}_{2} = \texttt{XOR}_{\thr}$. 

Partitioning parity blocks on the left and right 
helps to streamline these operations to optimize throughput.
This way, in one array, three gate operations can be active 
in each row at a time, as opposed to a mere one.
We keep multiple partitions (i.e., blocks) in the parity columns on each side.
    On each side, parity computations start in the left- or right-most blocks, respectively. After each step in computation, the next block closer to the computation columns gets used, moving one block at a time.
Due to parity block partitioning, each step of computation requires one parity block.

\vspace{0.02cm}
\noindent{\bf Generating Hamming Codes in Memory:} 

{A Hamming code with parameters $n$ and $k$, denoted as $Hamming(n,k)$, has
$n$ bit codewords with 
$n-k$ redundant bits (corresponding to check symbols) to enable error detection and correction, which we can think of as the equivalent of ``parity bits''.}
Specifically, in Equation~\eqref{eq:hamming}, the $k^{th}$ row of $A^T$ indicates which parity bits need to be updated when the $k^{th}$ bit in the original data vector is updated (as a result of computation). Since $A^T$ is a constant {$k\times(n-k)$} matrix, the parity update operation per bit in the original data reduces to up to $n-k$ \xor\ operations. 
Hence, all we need to generate Hamming codes is to update $n-k$ parity bits per operation instead of only one parity bit,
using the very same semantics and the data layout as in the previous discussion.
Since Hamming code has the equivalent of $n-k$ parity bits, 
once the second output of the \nortwo\ 
is produced in the left or right parity blocks, up to $n-k$ \xor\ computations follow.
And as \xor\ is implemented as a two-step operation, this requires up to $2(n-k)
$ gate operations. 

\subsection{Metadata Updates in \tmr}

\noindent \tmr's metadata corresponds to the two redundant copies of the computation results. 
Using standard, single output gates, we can generate these copies in the same row in a partitioned fashion, similar to the parity generation for \hamming,
\emph{
by implementing the same gate in three different column partitions simultaneously.}
At the same time, we can make use of multiple output gates that the targeted PiM technologies support, to generate the redundant copies in one shot, simultaneous with the actual computation.
{This can be achieved by using 3-output gates, in the same way as the 2-output gates for \hamming\ parity updates (Section~\ref{subsec:parity_pipeline}). We cover the electrical characterization for multiple output gates in the Appendix. Multiple-output gates have a higher power consumption when compared to their single-output counterparts,
which grows linearly with the number of outputs.}

\vspace{-.1cm}
\subsection{Single Error Protection Guarantee}
\label{sec:singleG}

\begin{figure}[t]
    \centering
\includegraphics[width=\linewidth,trim={0cm 0cm 0cm 0cm},clip]{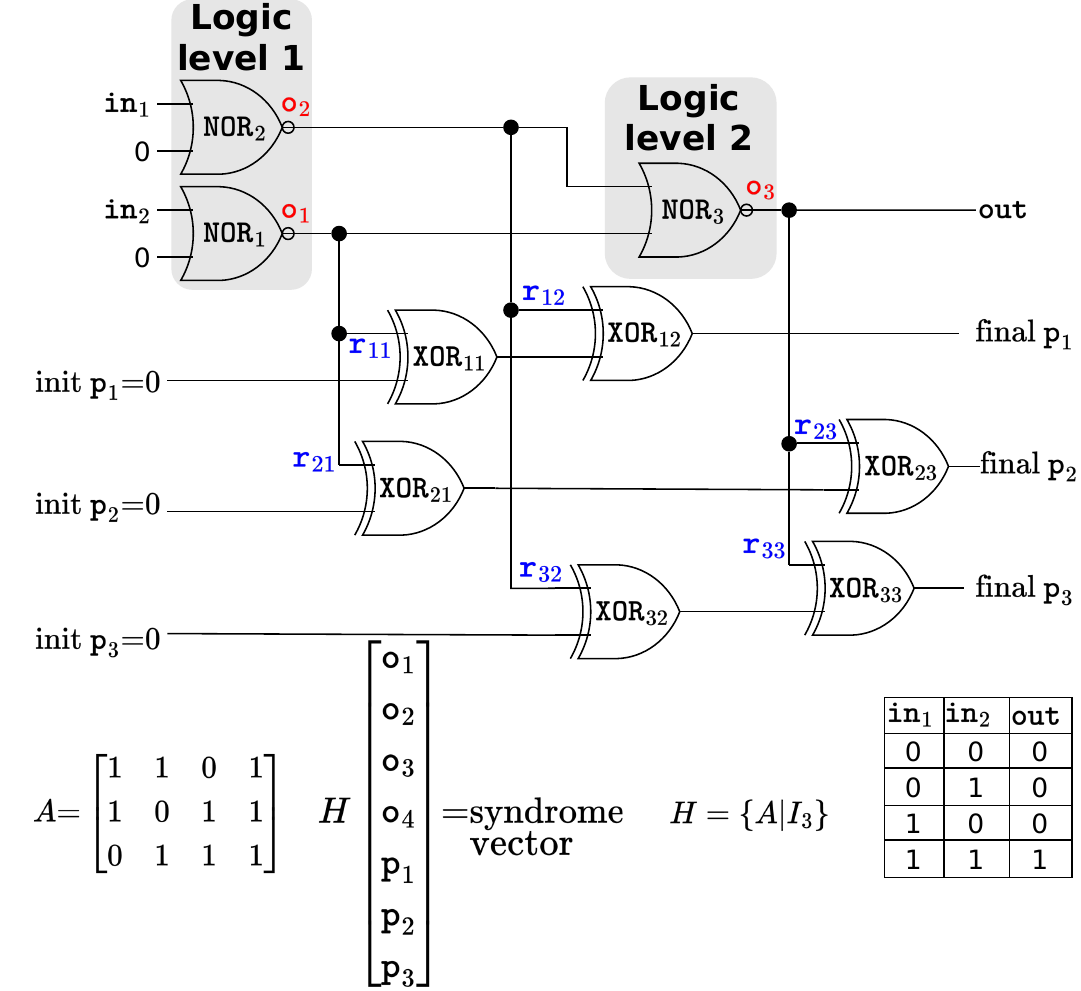}\\
\vspace{.1cm}
\scriptsize
\begin{tabular}{|c|c|c|c|}
\cline{2-3}
\multicolumn{1}{c|}{} & \multicolumn{2}
{|c|}{{\bf Number of Errors in}} & 
\multicolumn{1}{|c}{}\\
 \hline
{\bf Error} & {\bf Logic Level} & {\bf Final} & {\bf Final Error} \\
{\bf Site} & {\bf Output} & {\bf Output} & {\bf Outcome} \\
\hline
\multirow{3}{*}{$\texttt{o}_1$ or $\texttt{o}_2$} & \multirow{3}{*}{1} & \multirow{3}{*}{3} & Error propagates to \texttt{out}($\texttt{o}_3$),  \\
& & & and two respective parity bits \\
& & & if not fixed after logic level 1\\
\hline
$\texttt{o}_3$ & 1 & 1 & Error in \texttt{out} \\
\hline
$\texttt{r}_{11}$ or $\texttt{r}_{12}$ & 1 & 1 & Error in $\texttt{p}_1$ \\
\hline
$\texttt{r}_{21}$ or $\texttt{r}_{23}$ & 1 & 1 & Error in $\texttt{p}_2$ \\
\hline
$\texttt{r}_{32}$ or $\texttt{r}_{33}$ & 1 & 1 & Error in $\texttt{p}_3$ \\
\hline
\end{tabular}
    \caption{
    {
    Single error protection (SEP) guarantee.
    $\texttt{p}_\texttt{1},\texttt{p}_\texttt{2},\texttt{p}_\texttt{3}$  correspond to parity bits; $\texttt{out}$, to the final result of computation. 
    One data bit ($\texttt{o}_\texttt{4}$) remains unused, as our small illustrative example does not match a standard size Hamming code.
    }
    }
    \label{fig:hamming_drawing}
    \vspace{-0.5cm}
\end{figure}

\noindent Just adapting Hamming codes or TMR does not suffice to guarantee single error correction. The granularity in space and time at which we perform error checks dictates to what extent we can 
prevent errors in intermediate results to propagate to final outputs. 
Fig.\ref{fig:hamming_drawing} provides an illustrative \hamming\ example using 
$Hamming(7,4)$.
In a conventional (non-PiM) setting, $Hamming(7,4)$ can correct single errors by construction. 
The actual computation in this example is a sequence of three multi-output \nor\ gates -- $\texttt{NOR}_\texttt{1}$, $\texttt{NOR}_\texttt{2}$, $\texttt{NOR}_\texttt{3}$ -- implementing an $\texttt{AND}$ gate. 
Each $\texttt{o}$ represents a gate output, be it intermediate or not,
from the main computation.
Each $\texttt{r}$, on the other hand, corresponds to one 
redundant \nor\ output
used for parity updates.
As detailed in the Appendix, 
such $\texttt{o}$ and $\texttt{r}$ for multi-output gates are independent. 
{Because each \xor\ gate processes one independent \texttt{r} input, we can guarantee that any error in a given \texttt{r}
affects only a single parity bit and does not spread to other parity bits.}
\texttt{A} matrix from Fig.\ref{fig:hamming_drawing} assigns parity bits to gate outputs (Section~\ref{sec:background:fault_tolerance}); e.g., $\texttt{p}_{\texttt{1}}$ protects $\texttt{o}_{\texttt{1}}$, $\texttt{o}_{\texttt{2}}$.
To be more specific, $\texttt{r}_{\texttt{ij}}$ corresponds to 
{$\texttt{i}^{\texttt{th}}$ parity bit and $\texttt{j}^{\texttt{th}}$ logic 
{gate ($\texttt{NOR}_\texttt{j}$)}}, where $\texttt{i}, \texttt{j} = 1, 2, 3$.
{Each \nor\ gate in this example has 3 outputs.}
Each \xor\ corresponds to a cascade of a 2-output \nor\ and a threshold gate. $\text{\xor}_\texttt{ij}$ updates the $\texttt{i}^{\texttt{th}}$ parity bit with $\texttt{r}_{\texttt{ij}}$.
{We should also note that, even though \xor\ is implemented by two gates, any single error in \xor\ -- be it in inputs or component gates -- can only cause one bit flip at the respective \xor\ output.}
The table in Fig.\ref{fig:hamming_drawing} covers all possible cases for single errors (bit flips), where
the (second) third column tabulates the resulting number of errors at the output of (the respective logic level in) the main computation.
The first logic level comprises $\texttt{NOR}_\texttt{1}$ and $\texttt{NOR}_\texttt{2}$.
An error in $\texttt{o}_\texttt{1}$ ($\texttt{o}_\texttt{2}$) would result in a single bit error at the output of the first logic level
and $\texttt{NOR}_\texttt{3}$, as well as
in the corresponding parity bits $\texttt{p}_{\texttt{1}}$, $\texttt{p}_{\texttt{2}}$ ($\texttt{p}_{\texttt{1}}$, $\texttt{p}_{\texttt{3}}$). 
Therefore, to ensure that the number of errors remain at most one at the time of each error check -- such that the {\em Checker} can guarantee perfect correction -- we perform error checks at logic level granularity. 

Since ECC is maintained row-wise, potentially correlated errors along the columns are not a concern -- they have independent ECC by construction.
Error correlation along a row, on the other hand, can be temporal or spatial. 
Spatial correlation implies errors in confined portions of a row or a partition, where one gate operation would be performed at a time.
Each cell (that serves as a gate output) is preset before performing a gate operation, and presets happen alongside error correction. {A preset may overwrite an erroneous cell value and mask the error. Otherwise, if the preset itself is erroneous, the corresponding error can be corrected as a logic error. 
Spatially correlated errors can result in multiple errors per logic level if the correlation distance is smaller than the row space allocated for a logic level.}  
Temporal correlation implies multiple errors happening back-to-back in time (not necessarily in the same cell), which can lead to multiple errors per logic level
if temporally correlated errors are in the same logic level. We can always use stronger codes to protect against multi-bit errors as shown in Fig.\ref{fig:bch}, which our pipeline supports off-the-shelf. 
Errors in the outputs of 
multi-output gates can only be correlated if a shared circuit parameter such as the gate voltage is erroneous -- which can be avoided by strictly using single-output gates. However, unshared device parameters represent major sources of errors, such as critical current \cite{sun2012process}, temperature stability factor \cite{xie2016fokker} or tunneling magnetoresistance ratio \cite{zhao2012failure}.

\begin{table*}[h]
\centering

\begin{tabular}{c|c|c|c|c|c|c}
      & Update      & Check       & SEP  &  Time  & Energy & Checker  \\ 
      & Granularity & Granularity &   Guarantee   &              &        & Metadata \\
     \hline
     \multirow{3}{*}{\tmr} & Gate & Gate & $\checkmark$ & ${3N}$ & $ {3N}$ & ${2N}$ \\\cline{2-7}
      & \textbf{Gate} & \textbf{Logic Level} & \textbf{$\checkmark$} & {3N}, but can be fully masked  & \textbf{${3N}$} & \textbf{${2N}$} \\\cline{2-7}
     \hline
     \multirow{3}{*}{\hamming}  & Gate & Gate &\multicolumn{4}{|c}{Reduces to \tmr} \\\cline{2-7}
      & \textbf{Gate} & \textbf{Logic Level}& \textbf{$\checkmark$} & \textbf{$N(1{+}\log{N})$} & 
      \textbf{$N(1{+}\log{N})$}
      & {$N\log{N}$} 
     \\\cline{2-7}
     \hline
\end{tabular}
\caption{{Single Error Protection (SEP) design space for 
protecting $N$ 
PiM gate outputs. As highlighted in the table, our \tmr\ and \hamming\ designs perform metadata updates at gate; error checks, at logic level granularity. }
}
\label{tab:comparison}
\end{table*}

\subsection{Putting It All Together: Error Correction Design Space}

\noindent Table~\ref{tab:comparison} provides a comparison for different \hamming\ and \tmr\ design points, {constrained with the same area budget as the  unprotected counterparts. \hamming\ and \tmr\ designs therefore need to reclaim area throughout the execution, which adds to the time and energy overhead.  The overhead of area reclaims strongly depends on the application, and therefore, asymptotic trends in Table~\ref{tab:comparison} only cover the overheads for metadata updates and maintenance.} 

{\em Update Granularity} and {\em Check Granularity} refer to the granularity of metadata updates and error checks, respectively. By definition, {\em Check Granularity} cannot be finer than {\em Update Granularity}.
We consider {\em gate} and {\em logic level}
granularity. 
{\em Gate} implies updating meta data and/or triggering checks in the {\em Checker} after performing each Boolean gate operation; {\em logic level}, after performing all Boolean gates in a logic level (which are not data-dependent by construction). 
A {\em Check Granularity} of {\em circuit} is also possible, by triggering checks after performing all logic levels in the computation, however, cannot guarantee single error protection -- irrespective of the underlying {\em Update Granularity}, even a single gate error can propagate to gates in subsequent logic levels to result in multiple errors before the check takes place.  
Aside from {\em Time}
and {\em Energy} overhead, we also report how much space the {\em Checker} has to allocate for metadata -- which also serves as a proxy for the communication overhead between PiM arrays and {\em Checker} blocks.
\tmr\ with {\em Check Granularity = Gate} and {\em Update Granularity = Gate} corresponds to classic triple modular redundancy in time, which incurs a time overhead of 3$\times$. Error checks (which boil down to majority logic calculation) happen after each gate operation. 
{Timing optimizations like overlapping a gate operation in one row with the checks for another row, while possible, are less effective in this case, as the latency of checks and single gate operations are not well balanced.} 
Under {\em Check Granularity = Logic Level} such overlapped execution becomes more effective, by overlapping checks in one row with the computation of a logic level in another row -- which typically features many more gates than one. Accordingly, with sufficiently large logic levels, it is possible to mask the 3$\times$ time overhead of classic TMR.  

\hamming\ with {\em Check Granularity = Gate} and {\em Update Granularity = Gate}; i.e., {\em Hamming(3,1)}, reduces to \tmr\ operating at the very same granularity. 
Under {\em Check Granularity = Logic Level}, time and energy overhead evolve with the {\em Checker} metadata overhead, which becomes a logarithmic function of the number of bits (i.e., gate outputs) to be protected.

 {Note that \cite{leitersdorf2021making} also proposes a TMR implementation for PiM induced errors.
 However, \cite{leitersdorf2021making} uses PiM arrays for error correction, which limits protection capabilities. \tmr, on the other hand, only performs redundant computations in PiM arrays, and uses an external {\em Checker} for error detection and correction. \tmr\ also features 
 several design optimizations to minimize latency compared to an iso-area unprotected baseline.
 }

\section{Experimental Setup}\label{sec:experimental_setup}
\noindent We use a behavioral simulator for functional validation, alongside a
cycle-accurate timing simulator which can extract energy consumption and latency 
using the technology parameters in Table \ref{tab:params}. The simulator decomposes fixed point integer arithmetic into Boolean arithmetic and manages scratch space using a greedy memory allocator, which 
{{reclaim}s 
cells (whose data is no longer needed) 
whenever the array runs out of available scratch space.}
We utilize NVSim  \cite{dong2012nvsim} to estimate the peripheral circuitry overhead
induced by
sense amplifiers, column decoders, predecoder, charge/precharge, and driving control lines. To estimate the Hamming decoder and majority voter overhead in hardware, we utilize NanGate 45nm open cell library \cite{nangate45} and OpenROAD flow \cite{ajayi2019toward}. 

\begin{table}[thp]
    \centering
    \scalebox{1}{
    \begin{tabular}{c|c|c|c}
         Parameter & STT & SOT/SHE & ReRAM \\
         \hline
         $R_{low}/R_{ON}/R_{P}$ ($K\Omega$) & 3.15 \cite{zabihi2018memory} & 253.97 \cite{chowdhury2019spintronic} & 10 \cite{truong2021racer}\\
         $R_{high}/R_{OFF}/R_{AP}$ ($K\Omega$) & 7.34 \cite{zabihi2018memory} & 507.94 \cite{chowdhury2019spintronic} & 1000 \cite{truong2021racer}\\
         $R_{SHE}$ ($K\Omega$) & - & 64 \cite{chowdhury2019spintronic} & - \\
         $I_C$ ($\mu A$) & 50 \cite{zabihi2018memory} & 3 \cite{chowdhury2019spintronic} & -\\
         $V_{OFF}$/$V_{ON}$ (V/V) & - & - & 0.3/-1.5 \cite{talati2016logic} \\
         $t_\text{switch}$ ($ns$) & 1 \cite{chowdhury2019spintronic} & 1 \cite{chowdhury2019spintronic} & 1.3
         \cite{talati2016logic} \\
         \texttt{NOR} Energy ($fJ$) & 10.5 & 2.45 & 
         19.68 \\
         \texttt{THR} Energy ($fJ$) & 11.2 & 1.31 & 
         20.99 
         \\
         Write Energy ($fJ$) & 1.03 & 0.01 & 23.8
    \end{tabular}
    }
    \caption{Technology parameters. {$R_{SHE}$ captures the resistance of the SHE channel; $t_\text{switch}$, the switching time (i.e., gate delay), respectively.}
    } 
    \label{tab:params}
\end{table}

We experiment with representative benchmarks of different scales: 
{fixed-point Dense Matrix Multiplication,}
{a Multi-layer Perceptron (MLP),
and a larger scale FFT.}  
 {In dense matrix multiplication, we experiment with 8x8 ({\em mm8}), 16x16 ({\em mm16}), 32x32 ({\em mm32}), and 64x64 ({\em mm64}) matrices.}
 We use a two layer perceptron with 64 hidden neurons to classify MNIST, 
 considering 1--4 bits of weight precision ({\em mnist1, mnist2, mnist3, mnist4}).
We also include a variant of the in-memory FFT implementation from~\cite{cilasun2020crafft} 
as a representative larger scale benchmark for sensitivity analysis. {We experiment with 8, 16, 32, and 64 element FFTs ({\em fft8, fft16, fft32, fft64}).}
{While FFT can be implemented by ordinary 
sparse matrix-vector multiplications, our FFT benchmark is optimized for performance through butterfly arithmetic with complex numbers}.
We 
map all benchmarks to the underlying PiM substrates by synthesizing a PiM gate schedule in space and time.
{
We use
a fleet of PiM arrays, 
responsible for performing bulk bitwise logic as well as for maintaining parity state for \hamming\ and redundant computations for \tmr.
{
Depending on the application/problem size, we experiment with various array sizes in order to maximize the utilization.} All benchmarks are mapped to (no more than 16) $256\times256$ {arrays} for the whole computation.
{In order to match array dimensions and maximize array interface utilization, we use $Hamming(255,247)$ code with $n=255$ and $k=247$.} }

\section{Evaluation}\label{sec:evaluation}
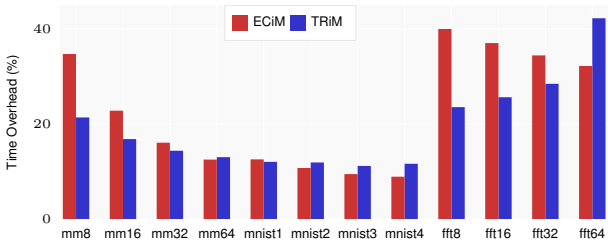
\begin{figure}
    \centering
\begin{tikzpicture}
\def\figFontSize{\sffamily\tiny}
\begin{axis}[
        ybar stacked,
        ylabel={Time Overhead (\%)},
        bar width=5pt,
        xtick={0,1,2,3,4,5,6,7,8,9,10,11},
        xticklabels={mm8, mm16, mm32, mm64, mnist1, mnist2, mnist3, mnist4, fft8, fft16, fft32, fft64},
        ymin=0,ymax=45,
        width=\linewidth,
        height=.5\linewidth,
        x label style={font=\figFontSize},
        y label style={font=\figFontSize}, 
        ticklabel style={font=\figFontSize},
        ylabel style={yshift=-.6cm},
        enlarge x limits=0.03,
        axis line style={draw=none},
        tick style={draw=none},
        legend entries={\hamming, \tmr},
        legend style={align=left, draw=none, font=\figFontSize, at={(0.42,1.00)},anchor=north, fill==gray!10, text opacity=1, draw=gray!20},
        legend cell align={left},
        legend image code/.code={%
            \draw[#1, draw=none] (0cm,-0.1cm) rectangle (0.2cm,0.1cm);
        },  
        legend columns=2,
        axis background/.style={fill=gray!5},
        grid=both, 
        grid style={white}
    ]
    \addplot +[bar shift=-2.5pt,draw=none,fill=red!60!gray] coordinates {(0, 34.71) (1, 22.79) (2, 16.06) (3, 12.51) (4, 12.55) (5, 10.74) (6, 9.46) (7, 8.91) (8, 39.97) (9, 37.01) (10, 34.42) (11, 32.19) };
    \resetstackedplots
    \addplot +[bar shift=2.5pt,draw=none,fill=blue!60!gray] coordinates {(0, 21.37) (1, 16.82)  (2, 14.37) (3, 13.02) (4, 12.03) (5, 11.90) (6, 11.17) (7, 11.63) (8, 23.54) (9, 25.62) (10, 28.45) (11, 42.23) }; 
\end{axis}
\end{tikzpicture}
\caption{Time overhead compared to an unprotected iso-area baseline. 
\vspace{-.5cm}
}\label{fig:latency}
\end{figure}

\begin{table}[t]
    \centering
    \scalebox{1.05}{
    \setlength\tabcolsep{1.5pt}
    \begin{tabular}{|c|c|c|c|c|c|c|c|c|c|c|c|c|}
\hline
\multirow{2}{*}{}& \multicolumn{4}{|c|}{mm} & \multicolumn{4}{|c|}{mnist} & \multicolumn{4}{|c|}{fft} \\
\cline{2-13}
& 8 & 16 & 32 & 64 & 1 & 2 & 3 & 4 & 8 & 16 & 32 & 64 \\
\hline
\hline
\hamming & 10 & 22 & 44 & 90 & 224 & 427 & 718 & 1162 & 26 & 42 & 60 & 87 \\
\hline
\tmr & 40 & 82 & 166 & 334 & 658 & 1414 & 2447 & 4295 & 118 & 241 & 445 & 1109 \\
\hline
    \end{tabular}
    }
    \caption{Number of area reclaims.}
    \label{tab:reclaims}
    \vspace{-.5cm}
\end{table}
\begin{table*}[t]
\centering
\begin{tabular}{|l|r|r|r|r|r|r|r|r|r|r|r|r|}
\cline{2-13}
\multicolumn{1}{c|}{\multirow{3}{*}{}}	& \multicolumn{6}{|c|}{\hamming} &  \multicolumn{6}{|c|}{\tmr} \\
\cline{2-13}
\multicolumn{1}{c|}{} & \multicolumn{2}{c|}{ReRAM} & \multicolumn{2}{|c|}{STT-MRAM} & \multicolumn{2}{|c|}{SOT-MRAM} &  \multicolumn{2}{c|}{ReRAM} & \multicolumn{2}{|c|}{STT-MRAM} & \multicolumn{2}{|c|}{SOT-MRAM} \\
\cline{2-13}
\multicolumn{1}{c|}{} & m-o & s-o & m-o & s-o & m-o	& s-o & m-o	& s-o & m-o	& s-o & m-o	& s-o \\
\hline
mm8	   & 19.55 & 81.23  & 43.26  & 355.03 & 360.56 & 1715.01 & \textbf{4.92}   & 17.98  & 6.42   & 46.50  & 37.67 & 175.82 \\
mm16   & 4.07  & 13.86  & 7.67   & 56.48  & 57.38  & 269.67  & \textbf{2.30}   & 6.44   & 2.23   & 11.30  & 7.80  & 33.43  \\
mm32   & 3.29  & 10.58  & 5.50   & 38.40  & 38.45  & 179.43  & \textbf{2.65}   & 7.91   & 2.34   & 12.19  & 6.54  & 27.40  \\
mm64   & \textbf{3.05}  & 9.51   & 4.40   & 29.23  & 28.47  & 131.90  & 3.42   & 11.05  & 2.76   & 15.66  & 5.89  & 24.30  \\
mnist1 & \textbf{32.97} & 122.27 & 31.38  & 247.01 & 207.62 & 984.11  & 53.89  & 201.84 & 37.37  & 295.90 & 36.84 & 171.57 \\
mnist2 & \textbf{39.20} & 151.33 & 28.90  & 229.16 & 167.46 & 793.48  & 87.04  & 339.91 & 57.87  & 466.58 & 38.43 & 179.22 \\
mnist3 & \textbf{46.30} & 184.38 & 28.96  & 231.31 & 149.64 & 708.98  & 117.75 & 474.00 & 76.55  & 624.02 & 39.47 & 184.26 \\
mnist4 & \textbf{56.70} & 230.97 & 31.55  & 253.94 & 147.60 & 699.49  & 167.31 & 688.08 & 107.25 & 881.51 & 46.06 & 215.71 \\
fft8   & 6.81  & 25.82  & 14.11  & 110.53 & 112.41 & 532.00  & \textbf{2.54}   & 7.58   & 2.78   & 15.93  & 12.41 & 55.41  \\
fft16  & 6.43  & 24.00  & 13.20  & 102.75 & 104.64 & 494.93  & \textbf{2.70}   & 8.20   & 2.96   & 17.38  & 13.30 & 59.63  \\
fft32  & 6.10  & 22.69  & 12.39  & 96.06  & 97.67  & 461.70  & \textbf{2.90}   & 9.11   & 3.19   & 19.32  & 14.48 & 65.27  \\
fft64  & 5.82  & 21.59  & 11.69  & 90.26  & 91.58  & 432.71  & \textbf{3.72}   & 12.63  & 4.20   & 27.79  & 20.37 & 93.34  \\
\hline
\end{tabular}
\caption{Energy overhead compared to an unprotected iso-area baseline. 
\emph{m-o}(\emph{s-o}) denote multi(single)-output gate designs. Highlighted are lowest overhead designs for each benchmark. 
\vspace{-.4cm}
}
\label{tab:energy}
\end{table*}

\noindent {In the following, we characterize the time and energy overhead of \hamming\ and \tmr\ under an {\em iso-area} budget when compared to an unprotected baseline for each benchmark. \hamming\ and \tmr\ designs therefore need to reclaim area throughout the execution, which adds to the time and energy overhead. While the area budget with fault tolerance is likely to be higher than the unprotected area budget, area increases to totally eliminate area reclaims may not always be feasible considering asymptotic trends and practical limitations (Table~\ref{tab:comparison}). In this regard, the reported overheads can be regarded as worst-case overheads.}

\noindent{\bf Time Overhead:} 
Fig.\ref{fig:latency} shows the time overhead with respect to an unprotected iso-area baseline, using 
multi-output gates.
For both, \hamming\ and \tmr, 
the main contributing factor to the latency is metadata updates 
along with area reclaims, 
i.e., the number of times 
used scratch space is recycled 
to meet the area budget.
Since \tmr\ metadata takes up notably more space than \hamming, in \tmr, less
scratch space is available for main computation.
As a consequence, larger problem sizes incur more area reclaims, and the time overhead for \tmr\ grows more significantly.
Overall, \tmr\ outperforms \hamming\ for smaller problem sizes. The opposite applies at scale. For the largest problem we considered ({\em fft64}), \hamming\ has a lower time overhead {({29\%})} than \tmr\ {({42\%})}. This trend (Fig.\ref{fig:latency}) is in line with the number of area reclaims from Table \ref{tab:reclaims}. With growing problem sizes the logarithmic latency overhead of \hamming\ (Table~\ref{tab:comparison}) gets amortized, while metadata overheads (including the data communication with the {\em Checker}) become even more significant for \tmr.

\noindent{\bf Energy Overhead:} 
Table \ref{tab:energy} summarizes the energy overhead compared to an iso-area unprotected baseline, considering different
technologies. We also make a distinction between \hamming\ and \tmr\ implementations using single-output and multi-output gates.
Overall, in smaller benchmarks, \hamming\ has a higher energy overhead due to parity updates 
being performed for every \texttt{NOR} operation in the actual computation. Still, depending on the application requirements and error characteristics, the efficiency in the area vs. latency trade-off can mask this energy drawback. 
Different benchmarks give rise to different circuit topologies, which dictate the number of logic levels (circuit depth). Logic level count tends to increase with problem size, and determines the frequency of checks, hence the communication overhead with the {\em Checker}.
For larger benchmarks, area reclaims and communication overheads become more significant, and more so for \tmr, as \tmr\ incurs more area reclaims and higher-volume data communication with the {\em Checker}.

The differences in the energy overhead of different technologies for the same design configuration mainly stem from the differences in the relative energy of gate operations with respect to writes. 
Generally, if error correction in memory is more computation heavy (due to more complex metadata updates as for \hamming) than data communication (due to larger volume data transfers to/from the {\em Checker} as for \tmr), relatively higher energy of writes with respect to gate operations can render lower energy overheads and vice versa, granted that a break even point exists.

The energy overheads in Table \ref{tab:energy} reflect 
practical factors such as 
the energy overhead of area reclaims or
data communication with the {\em Checker} as well as the {\em Checker} energy, which are not considered in 
Table \ref{tab:comparison}. 

\noindent{{{\bf Extension to Higher-Coverage Codes:} Beyond Hamming codes,
\hamming\
can also support a class of codes that can correct multiple errors. One example is BCH codes \cite{bose1960class,hocquenghem1959codes}, where multiple error correction is possible at the expense of an increased number of parity bits, as captured by Fig.\ref{fig:bch}. We use $Hamming(255,247)$ by default with $n=255$ and $k=247$. BCH codes are characterized by the same parameters, and we sweep $k$ for a fixed $n$ in this analysis.
Similar to Hamming codes, BCH codes can be updated in memory just by updating the corresponding parity bits from the non-identity part of the generator matrix $G$ (\emph{i.e.}, $-A^T$)}} and can be implemented following the exact same principles outlined in Section~\ref{subsec:parity_pipeline}. 
{Latency and energy overhead of \hamming\ is {proportional to the number of parity bits to be maintained}, which grows in a sublinear fashion in the more general case of BCH codes. } 

\begin{figure}[h]
\vspace{-.2cm}
    \centering
    \includegraphics[trim={.22cm .065cm .59cm .26cm},clip,width=.55\linewidth]{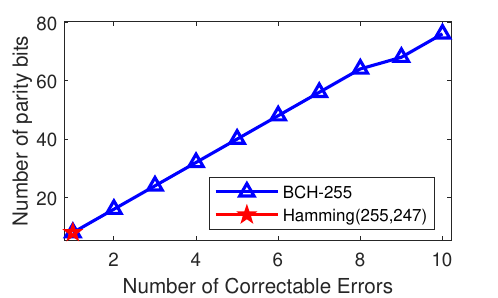}
    \caption{{Number of parity bits vs. correctable errors.
    }
    }
    \label{fig:bch}
    \vspace{-.3cm}
\end{figure}

\section{Related Work}\label{sec:related_work}
\noindent Homomorphic codes are appealing candidates for ECC in PiM applications, since extracting output parity solely using input parities 
maximizes
efficiency. For example, Reed-Muller codes support additive and multiplicative homomorphism and are promising for homomorphic computing \cite{cho2020homomorphic}. However, addition in original data becomes XOR-addition in codewords, while multiplication becomes even-costlier cyclic convolution. Simple operations on original data translate into orders-of-magnitude costlier (codeword) computations, resulting in excessive overhead for bulk bitwise operations. 
Other linear block codes 
don't
support binary element-wise multiplication cost-efficiently\cite{han2017homomorphic}, 
required to implement universal Boolean logic.
Arithmetic codes such as Berger, Residue, AN, ANB, and ANBD inherently support homomorphism for addition and multiplication \cite{schiffel2011hardware}. Only Berger codes allow direct computation of bitwise operations homomorphically, but still not cost-effectively, as output check symbols don't only depend on input check symbols. 

Reliable-Simpler-MAGIC supports general-purpose in-memory ECC, capable of detecting and correcting only storage related errors, and not logic errors induced by computations in memory \cite{leitersdorf2021efficient,kvatinsky2021making}. 
The idea is diagonally calculating parities once the data is written into the PiM substrate,
before and after sensitive tasks. 
The only relevant work which can correct general-purpose PiM logic errors is the TMR approach in \cite{leitersdorf2021making}, which (with several extra optimizations
to maximize fairness) 
we consider
in our evaluation.
This paper does not specify how to perform majority voting, data transfers to/from the voting module, or operation under memory size constraints. We cover all of these aspects in our MR-based solution \tmr.
More precisely: (1) We introduce an external checker module to guarantee correct operation, while majority voting for TMR is not protected in \cite{leitersdorf2021making}. (2)
We propose a pipeline 
to mask data communication overheads with the external checker at logic level granularity. 
(3) We provide a detailed quantitative characterization of time and energy overheads under strict area constraints, which is not provided in \cite{leitersdorf2021making}.
Moreover, \tmr\  is just one of the designs we propose in our study. We also introduce a PiM-specific Hamming code based design (\hamming) of minimal time overhead under strict area budget constraints. 

{More recent work \cite{li2022error} 
on ECC for PiM with memristive devices uses four extra gate operations to protect a single Boolean gate computation, incurring a significantly higher overhead than 
our solutions
by construction.} 
{Another recent study, FAT-PiM \cite{fatPiM} 
focuses on only detection in crossbars.
FAT-PiM is based on a summation idea that 
cannot accommodate
general-purpose bulk bitwise computation in PiM technologies we consider. 
This also is the case for \cite{feinberg2018making}, which uses arithmetic codes in a dot-product crossbar setting
specialized for matrix-vector multiplication.
Arithmetic (more specifically, AN) codes can efficiently protect linear algebraic computations, but not bitwise logic \cite{schiffel2011hardware}.
Only Berger codes support bitwise logic, which are not feasible for PiM due to complex data dependencies. In contrast, our paper targets general purpose computations in memory via bulk bitwise logic -- capable of performing matrix vector multiplication, but not limited to linear algebraic computations 
due to the support for universal logic gates. 
Nevertheless, being tightly tailored to an application domain, designs like \cite{feinberg2018making} incur less overhead than our general purpose solutions.  
Finally, while only detection can still be acceptable
for applications with a sufficient budget for recomputation upon error detection, 
direct correction is desirable for wider spread adoption.}

\section{Conclusion \& Discussion}\label{sec:conclusion}
\noindent While memory errors are extensively studied, error detection and correction for processing in memory (PiM) requires
rethinking
due to the highly dynamic nature of the data to be protected.  In this study, we investigate various techniques to improve the reliability of nonvolatile PiM operations. We compile 
the specification for a PiM-oriented ideal error correction scheme and explore the design space.

We propose several solutions based on Hamming codes (\hamming) and TMR ({\tmr}),
and characterize their time and space complexity along with energy efficiency under iso-error-coverage (guaranteed single error correction),
considering representative nonvolatile PiM technologies -- spanning ReRAM and MRAM variants. 
Key novel aspects which apply to all design points include: Single error protection guarantee due to error correction and detection at {\em logic level} granularity; full system design featuring an external checker with optimized data transfer to/from the checker; design modularity and straight-forward extension to stronger codes with protection guarantees for larger number of errors (such as BCH).
\hamming\ and \tmr\ by construction provide {guaranteed} protection against computation-induced errors and inherently cover memory/storage errors, including 
potential errors in the input data.

We assume that the PiM substrate corresponds to a stand-alone accelerator, which may be connected to a host machine but which is not tightly integrated into the memory hierarchy of the host. In this case we can provide full coverage for both logic and memory. We can catch memory errors if corresponding memory cells serve as gate inputs or outputs.
When used as a dedicated accelerator for large scale memory intensive applications, this generally is the case for the vast majority of the cells \cite{nejatollahi2020cryptopim, cilasun2021spiking, imani2019floatpim, chowdhury2022cram, khalifa2021filtpim}. 

However, even if memory cells do not serve as logic gate inputs or outputs, we can designate the ``computation'' as a trivial buffering operation to cover pure memory errors. 

If, on the other hand, the PiM substrate is tightly integrated with the memory hierarchy and is only occasionally used for smaller scale computations (as opposed to a dedicated accelerator), it is plausible to assume that the portion of the memory contained in the PiM substrate remains unchanged most of the time. Accordingly, lower-overhead classical fault tolerance techniques may become applicable.

\section*{Appendix: Electrical Characterization}
\label{subsec:ec}
\begin{figure}[tph]
\centering
\subfloat[]{\label{fig:noisemargin}{\includegraphics[width=.48\linewidth]{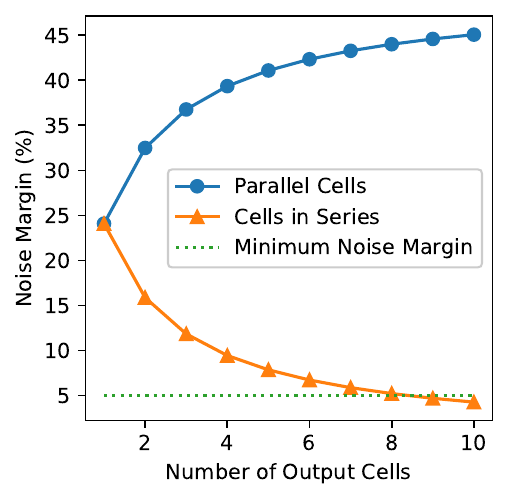}}}
\subfloat[]{\label{fig:voltages}{\includegraphics[width=.5\linewidth]{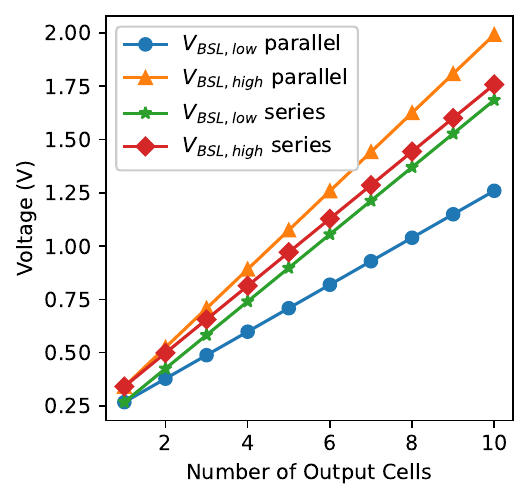}}}
\caption{
Noise margins (a) and 
voltages (b).
}
\label{fig:noisemargin_voltages}
\vspace{-.3cm}
\end{figure}

{
\noindent Correct operation 
relies on the following conditions:}
\begin{list}{\labelitemi}{\leftmargin=.5em}
 \itemsep 0.5em 
    \item [{\bf 1)}]
    {\textbf{Support for 2-output gates:} The 2-step \xor\ operation requires 2-output \nor\ (\nortwo) gates. }
    \item [{\bf 2)}]
    {\textbf{Bias voltage matching:} Since multiple partitions in the same array are controlled using the same column control lines, i.e., 
    WLs and WLW/Rs, they need to operate under the same bias voltages. Therefore, we need to ensure that different types of logic components in the implementation, 
    i.e., 
    \nortwo\ and 4-input \thr\ gates, work within the same bias voltage range.}
\end{list}

\noindent{\bf STT and SOT/SHE:}\label{sec:detection_correction:sttshe_multi_output}
Extending the analysis in \cite{zabihi2018memory} to the serial and parallel (with respect to how the resistances of the two outputs are connected in the resistive division network corresponding to the underlying logic gate) multiple output gate operation, using the  ``Today's MTJ'' parameter set (of \cite{zabihi2018memory}), we obtain Equation \eqref{eq:parallel} for the parallel case:
\begin{equation}\label{eq:parallel}
\footnotesize
\begin{aligned} 
V_{BSL,low (\text{parallel})} = N{\frac{(TMR+1)R_P}{TMR+2} + \frac{R_P}{N}}{I_C},\\
V_{BSL,high (\text{parallel})} = N{\frac{(TMR+1)R_P}{2} + \frac{R_P}{N}}{I_C}
\end{aligned}
\end{equation}
Equation \eqref{eq:series} captures the serial case:
\begin{equation} \label{eq:series}
\footnotesize
\begin{aligned} 
V_{BSL,low (\text{series})} = {\frac{(TMR+1)R_P}{TMR+2} + R_P N}{I_C},\\
V_{BSL,high (\text{series})} = {\frac{(TMR+1)R_P}{2} + R_P N}{I_C}
\end{aligned}
\end{equation}
{The low and high voltages are obtained by 
Kirchoff's laws on the equivalent resistances for marginally non-switching and switching cases in the truth table.}
Here, TMR is the Tunnel Magneto-Resistance; $R_P$, the MTJ parallel state (i.e., low) resistance;  $I_C$, the critical switching current; and $N$, the number of output cells. 
{Noise margin (\%) is defined as
$\frac{V_\text{high}-V_\text{low}}{\frac{V_\text{high}+V_\text{low}}{2}}$ in \cite{zabihi2018memory} 
as a measure of error tolerance of the gate implementation.} We show the resulting noise margins for the multiple-output case, considering both serial and parallel connectivity, in Fig.\ref{fig:noisemargin}.
We assume 
{a $5\%$} {minimum} {noise margin}. 
The corresponding bias voltages are  provided in Fig.\ref{fig:voltages}. 
{
This analysis reveals that 
multi-output implementations are feasible
and more efficient when the output MTJs are placed in parallel.}
{Further, {\thr}'s bias voltage ($V_{bias}$) should satisfy:}
\begin{equation} \label{eq:STT_THR_V_bias}
\footnotesize
\begin{aligned}
I_C \left( R_P \parallel R_P \parallel R_P \parallel R_{AP} + R_P\right) < V_{bias} < \\
I_C \left( R_P \parallel R_P \parallel R_{AP} \parallel R_{AP} + R_P\right)
\end{aligned}
\end{equation}
{To ensure that both \thr\ and \nor\ 
correctly work in the same bias voltage range, we introduce \texttt{D} dummy inputs to \nor\ gates.}
{
\texttt{D} is 4 for STT; 
5, for SOT/SHE; and 2,
for ReRAM.}
{For an \texttt{N}-output \texttt{NOR} with \texttt{D} 
dummy 
inputs to match the voltage range of the thresholding gate, we can characterize the voltage requirements as follows:}
\begin{equation} \label{eq:STT_NOR_V_bias}
\footnotesize
\begin{aligned}
N I_C \left( R_P \parallel R_P \parallel \frac{R_P}{D} + \frac{R_P}{N}\right) < V_{bias} < \\
N I_C \left( R_P \parallel R_{AP} \parallel \frac{R_P}{D} + \frac{R_P}{N}\right)
\end{aligned}
\end{equation}
{\texttt{D} depends on the technology parameters, and can be easily tuned to find an overlapping bias voltage range within which both  
the \nor\ and the \thr\
gates can operate. In the SHE case, besides the changes in the technology parameters, the only difference in Eqs. \ref{eq:STT_THR_V_bias} and \ref{eq:STT_NOR_V_bias} is the output resistance, which now becomes
the resistance of the SHE channel, respectively. 
{Note that here $\frac{R_P}{N}$ is the output resistance and the remaining resistance values in the parenthesis correspond to the marginal input combination.} Our observations so far hold otherwise.}

\noindent{\bf ReRAM:}
\label{reram_multi_output}
{
The low and high resistance states are characterized by $R_{ON}$ and $R_{OFF}$; threshold voltages, by 
$V_{OFF}$ and $V_{ON}$.
We can derive \thr\ bias voltage 
from}
\begin{equation} \label{eq:ReRAM_THR_V_bias}
\footnotesize
\begin{aligned}
\frac{V_{OFF}}{R_{ON}} \left( R_{ON} + R_{OFF} \parallel R_{OFF} \parallel R_{ON} \parallel R_{ON} \right) < V_{bias} < \\
\frac{V_{OFF}}{R_{ON}}\left(R_{ON} + R_{OFF} \parallel R_{OFF} \parallel R_{OFF} \parallel R_{ON} \right)
\end{aligned}
\end{equation}
and the \nortwo\ bias voltage 
from
\begin{equation} 
\label{eq:ReRAM_NOR_V_bias}
\footnotesize
\begin{aligned}
\frac{V_{OFF}}{\frac{R_{ON}}{N}} \left( {\frac{R_{ON}}{N}} + R_{OFF} \parallel R_{ON} \parallel \frac{R_{ON}}{D}\right) < V_{bias} < \\
\frac{V_{OFF}}{\frac{R_{ON}}{N}}\left({\frac{R_{ON}}{N}} + R_{OFF} \parallel R_{OFF} \parallel \frac{R_{ON}}{D}\right)
\end{aligned}
\end{equation}

{
\noindent We can match the \nor\ and \thr\ bias voltage range 
in a similar manner to STT or SOT\slash SHE otherwise.
}

\bibliographystyle{IEEEtranS}
\bibliography{refs,wangrefs}

\clearpage
\thispagestyle{empty}

\end{document}